\documentclass[11pt,fleqn]{article}

\usepackage{amsmath,amsfonts,amssymb,latexsym,cite}
\usepackage{graphicx}

\usepackage{amsfonts,amssymb,cite}
\usepackage{graphicx}

\def\noi{\noindent}

\newcommand{\Title}[1]{\noi {{\Large\bf #1}}\\[1ex]}

\def\Aunames#1{\noi{\bf #1}}
\def\au#1{${}^{#1}$}
\def\Addresses#1{\medskip\noi \protect
	\begin{description}\itemsep -3pt {\it #1} \end{description}}
\def\adr#1#2{\item[${}^{#1}$]{\it #2}}

\newcommand{\Abstract}[1]{\vskip 2mm \begin{center}
        \parbox{16.4cm}{\small\noi #1} \end{center}\medskip}

\def\email#1#2{\footnotetext[#1]{e-mail: #2}\addtocounter{footnote}{1}}

\def\nqq{\hspace*{-2em}}

\usepackage{color}

\def\Jl#1#2{#1 {\bf #2},\ }

\def\ApJ#1 {\Jl{Astroph. J.}{#1}}
\def\CQG#1 {\Jl{Class. Quantum Grav.}{#1}}
\def\DAN#1 {\Jl{Dokl. AN SSSR}{#1}}
\def\GC#1 {\Jl{Grav. Cosmol.}{#1}}
\def\GRG#1 {\Jl{Gen. Rel. Grav.}{#1}}
\def\IJMPD#1 {\Jl{Int. J. Mod. Phys. D}{#1}}
\def\JETF#1 {\Jl{Zh. Eksp. Teor. Fiz.}{#1}}
\def\JETP#1 {\Jl{Sov. Phys. JETP}{#1}}
\def\JHEP#1 {\Jl{JHEP}{#1}}
\def\JMP#1 {\Jl{J. Math. Phys.}{#1}}
\def\NPB#1 {\Jl{Nucl. Phys. B}{#1}}
\def\NP#1 {\Jl{Nucl. Phys.}{#1}}
\def\PLA#1 {\Jl{Phys. Lett. A}{#1}}
\def\PLB#1 {\Jl{Phys. Lett. B}{#1}}
\def\PRD#1 {\Jl{Phys. Rev. D}{#1}}
\def\PRL#1 {\Jl{Phys. Rev. Lett.}{#1}}

\def\lal{&&\nqq {}}

\def\beq{\begin{equation}}
\def\eeq{\end{equation}}
\def\bear{\begin{eqnarray}}
\def\bearr{\begin{eqnarray} \lal}
\def\ear{\end{eqnarray}}
\def\earn{\nonumber \end{eqnarray}}

\def\nnn{\nonumber\\ \lal }

\def\yy{\\[5pt] {}}
\def\yyy{\\[5pt] \lal }

 \catcode`\@=11 \@addtoreset{equation}{section}\catcode`\@=12

\begin{document}
\twocolumn[
%\jnumber{4}{2019}

\Title{Dyon-like black hole solutions in the model 
       with two Abelian gauge fields  }

\Aunames{ M. E. Abishev\au{a,1}, V. D. Ivashchuk\au{b,c,2}, A. N. Malybayev\au{a} and S. Toktarbay\au{a}} 

\Addresses{
\adr a {\small 
    Institute of Experimental and Theoretical Physics, 
    Al-Farabi Kazakh National University,  \yy
    Al-Farabi avenue, 71, Almaty 050040, Kazakhstan }
\adr b {\small Peoples' Friendship University of Russia (RUDN University), 
             ul. Miklukho-Maklaya 6, Moscow 117198, Russia}          
\adr c {\small Center for Gravitation and Fundamental Metrology, 
  %VNIIMS,
  ul. Ozyornaya  46, Moscow 119361, Russia}
	}

%\Dates{August 8, 2019}{August 10, 2019}{August 22, 2019}

\Abstract
 {Dilatonic black hole dyon-like solutions in the gravitational
  $4d$ model with a scalar field,  two 2-forms, two dilatonic coupling constants
  $\lambda_i \neq 0$, $i =1,2$, obeying $\lambda_1 \neq - \lambda_2$
  and the sign parameter $\varepsilon = \pm 1$ for scalar field kinetic term
  are overviewed. Here $\varepsilon = - 1$ corresponds to a phantom scalar field. 
  The solutions are defined up to solutions of two master equations for two moduli functions, when
   $\lambda^2_i \neq 1/2$ for $\varepsilon = - 1$. Several integrable cases
   corresponding  to  Lie algebras $A_1 + A_1$, $A_2$, $B_2 = C_2$ and $G_2$ are considered.
  Some physical parameters of the solutions are derived:
  gravitational mass, scalar charge, Hawking temperature, black hole area entropy and
  PPN parameters $\beta$ and $\gamma$. Bounds on the gravitational mass
  and scalar charge (based on a certain conjecture)  are presented. }

] %%%%%%%%%%%%%%%%%%%%%%%
\email 1 {abishevme@mail.ru}
\email 2 {ivashchuk@mail.ru}

{ % au-def
\def\R{{\mathbb R}}

% ====================================
\section{Introduction}

 At present there exists a certain interest in spherically symmetric solutions,
 e.g. black hole and black brane ones, related to Lie algebras and
Toda chains, see \cite{BronShikin}-\cite{GalZad}
and the references therein. These solutions appear
in gravitational models with scalar fields and antisymmetric forms.

Here we overview dilatonic black hole solutions with  electric and magnetic
charges $Q_1$ and $Q_2$, respectively, in the $4d$ model with metric $g$,  scalar field $\varphi$,  two 2-forms  $F^{(1)}$ and $F^{(2)}$, corresponding to two dilatonic coupling constants $\lambda_1$ and $\lambda_2$, respectively.
All fields are defined on an oriented manifold ${\cal M }$. 
Here we  deal with the dyon-like configuration for fields of 2-forms:
\beq \label{0.1}
 F^{(1)} = Q_1 e^{2 \lambda_1 \varphi} * \tau, \qquad F^{(2)} = Q_2 \tau,
 \eeq
where $\tau = {\rm vol}[S^2]$ is volume form on $2d$ sphere and $* = *[g]$ is the Hodge operator
corresponding to the oriented manifold ${\cal M }$  with the metric $g$. 
The ansatz (\ref{0.1}) means that we deal here with a charged black hole, which has two color charges:
$Q_1$ and $Q_2$. The charge $Q_1$ is the electric one corresponding to the form $F^{(1)}$, while 
the charge $Q_2$ is the magnetic one corresponding to the form $F^{(2)}$. For coinciding dilatonic couplings 
$\lambda_1 = \lambda_2 = \lambda$ we get a trivial noncomposite generalization of 
dilatonic dyon black hole solutions in the model with one 2-form which was considered in ref. \cite{ABDI}, see also 
\cite{Lee,ChHsuL,GKLTT,PTW,FIMS,GKO} and references therein.

 The main motivation for considering this and more general $4D$ models 
governed by the Lagrangian  density ${\cal  L}$:
 \bearr \label{i.16r}
  {\cal  L}/ \sqrt{|g|} =  
   R[g] -   h_{ab} g^{\mu \nu} \partial_{\mu} \varphi^a  \partial_{\nu} \varphi^b \nnn
    - \frac{1}{2} \sum_{i =1}^{m} \exp(2 \lambda_{i a}  \varphi^{a})F^{(i)}_{\mu \nu} F^{(i)\mu \nu},      
  \ear
where $\varphi =(\varphi^a)$ is a set of $l$ scalar fields,  $F^{(i)} = d A^{(i)}$
are 2 forms and  $\lambda_{i} = (\lambda_{i a})$ are dilatonic coupling vectors,
$i =1, \dots, m$, is coming from dimensional reduction of supergravity models;
in this case the matrix  $(h_{ab})$ is positive definite.  For example, one may consider 
a part of bosonic sector of  dimensionally reduced  $11d$ supergravity \cite{LP} 
with $l$ dilatonic scalar fields and $m$ $2$-forms (either  originating  from 11d metric  
or  coming from $4$-form) activated; Chern-Simons terms  vanish in this case.  
Certain uplifts (to higher dimensions) of 4d black hole solutions corresponding to  
(\ref{i.16r}) may lead us to black brane solutions in dimensions $D >4$, 
e.g. to dyonic ones; see \cite{LP,DLP,IMp2,CGLO,GO} and the
references therein. The dimensional reduction from the 12-dimensional model from 
ref. \cite{KKLP} with phantom scalar field and two forms of rank $4$ and $5$ will 
lead us to the Lagrangian  density
(\ref{i.16r}) with the matrix $(h_{ab})$ of pseudo-Euclidean signature. 

The dilatonic scalar field may be either an ordinary one or a phantom (or ghost) one.
The phantom field   appears in the action with a kinetic term of the
``wrong sign'', which implies the violation of the null energy condition $p  \geq - \rho
$. According to ref. \cite{ArkH}, at the quantum level, such fields could form a
``ghost condensate'', which may be responsible for modified
gravity laws in the infra-red limit. The observational data do not exclude this possibility \cite{Kom}.

Here we present certain  relations for the physical parameters of dyonic-like black holes, e.g.
bounds on the gravitational mass $M$ and the scalar charge $Q_{\varphi}$. As in our previous work \cite{ABDI}
 this problem is  solved here up to a conjecture, which states a one-to-one (smooth) correspondence between
the pair  $(Q_1^2,Q_2^2)$, where $Q_1$ is the electric charge and $Q_2$ is the magnetic charge,
and the pair of positive parameters $(P_1,P_2)$, which appear in decomposition of
 moduli functions at large distances. Here we use analogous
 conjecture which is believed to be valid for all $\lambda_i \neq 0$
in the case of an ordinary scalar field and for $0< \lambda_i^2 < 1/2$ for the case of a phantom scalar field 
(in both cases the inequality $\lambda_1 \neq - \lambda_2$ is assumed).

% ====================================
\section{Black hole dyon solutions}

We satrt with a model governed by the action
\bearr  
 S= \frac{1}{16 \pi G}  \int d^4 x \sqrt{|g|}\biggl\{ R[g] -
 \varepsilon g^{\mu \nu}\partial_{\mu} \varphi  \partial_{\nu} \varphi
 \qquad \qquad   \nnn
 - \frac{1}{2} e^{2 \lambda_1 \varphi} F^{(1)}_{\mu \nu} F^{(1)\mu \nu }
 - \frac{1}{2} e^{2 \lambda_2 \varphi} F^{(2)}_{\mu \nu} F^{(2) \mu \nu}
 \biggr\} \label{i.1},
\ear
where $g= g_{\mu \nu}(x)dx^{\mu} \otimes dx^{\nu}$ is  metric,
 $\varphi$ is the scalar field, $F^{(i)} = dA^{(i)}
 =  \frac{1}{2} F^{(i)}_{\mu \nu} dx^{\mu} \wedge dx^{\nu}$
is the $2$-form with $A^{(i)} = A^{(i)}_{\mu} dx^{\mu}$, $i =1,2$, $\varepsilon = \pm 1$,
$G$ is the gravitational constant,
 $\lambda_1, \lambda_2 \neq 0$ are  coupling constants obeying $\lambda_1 \neq - \lambda_2$ and
 $|g| =   |\det (g_{\mu \nu})|$. Here we also put $\lambda_i^2 \neq 1/2$, $i =1,2$,
 for $\varepsilon = - 1$. 
 
 We consider a family of dyonic-like black hole
solutions to the field equations corresponding to the action
(\ref{i.1}) which are defined on the manifold
\beq \label{i.2}
 {\cal M }  =    (2\mu, + \infty)  \times S^2 \times  \R,
\eeq
and have the following form
\bearr  \label{i.3}
 ds^2 = g_{\mu \nu} dx^{\mu} dx^{\nu}
 \yyy
 = H_1^{h_1} H_2^{h_2}
 \biggl\{ -  H_1^{-2 h_1} H_2^{-2 h_2}
 \left( 1 - \frac{2\mu}{R} \right)
 dt^2 \nnn
  \qquad +  \frac{dR^2}{1 - \frac{2\mu}{R}} + R^2  d \Omega^2_{2}
  \biggr\}, 
 \nnn  
 \label{i.3a}
 \exp(\varphi)=
 H_1^{h_1 \lambda_1 \varepsilon} H_2^{- h_2 \lambda_2 \varepsilon },
 \yyy  \label{i.3be}
 F^{(1)}=
 \frac{Q_1}{R^2}   H_{1}^{-2} H_{2}^{- A_{12}} dt \wedge dR,
  \yyy  \label{i.3bm}
 F^{(2)}  =  Q_2 \tau.
\ear
Here  $Q_1$ and $Q_2$ are (colored) charges -- electric and magnetic, respectively,
 $\mu > 0$ is the extremality parameter,
 $d \Omega^2_{2} = d \theta^2 + \sin^2 \theta d \phi^2$
is the canonical metric on the unit sphere $S^2$
 ($0< \theta < \pi$, $0< \phi < 2 \pi$),
 $\tau = \sin \theta d \theta \wedge d \phi$
is the standard volume form on $S^2$,
\beq \label{i.16}
 h_i = K_i^{-1}, \qquad K_i = \frac{1}{2} +  \varepsilon \lambda_i^2,
\eeq
$i =1,2$, and
\beq \label{i.16a}
  A_{12} =  (1 - 2 \lambda_1 \lambda_2 \varepsilon) h_2.
\eeq
The functions $H_s > 0$ obey the equations
\beq \label{i3.1}
 R^2 \frac{d}{dR} \left( R^2
  \frac{\left(1 - \frac{2 \mu}{R}\right)}{H_s}
  \frac{d H_s}{dR} \right) = - K_s  Q_s^2
  \prod_{l = 1,2}  H_{l}^{- A_{s l}},
\eeq
with the following boundary conditions imposed:
\beq \label{i3.1a}
  H_s  \to H_{s0} > 0
\eeq
for $R \to 2\mu $, and
\beq \label{i3.1b}
  H_s    \to 1
\eeq
for $R \to +\infty$, $s = 1,2$.

In (\ref{i3.1}) we denote
\beq \label{i.18}
    \left(A_{ss'}\right)=
  \left( \begin{array}{*{6}{c}}
     2 &  A_{12}\\
     A_{21} & 2\\
\end{array}
\right) ,
\eeq
where $A_{12}$ is defined in (\ref{i.16a}) and
\beq \label{i.16b}
  A_{21} =  (1 - 2 \lambda_1 \lambda_2 \varepsilon) h_1.
\eeq

These solutions  may be obtained just by using  general formulas for non-extremal (intersecting)  black brane solutions from \cite{IMp1,IMp2,IMp3} (for a review see \cite{IMtop}).
The composite analogs of the solutions  with one 2-form and $\lambda_1 = \lambda_2$
 were presented  in ref. \cite{ABDI}.

The first boundary condition (\ref{i3.1a}) guarantees (up to a possible additional requirement on the analyticity 
of $H_s(R)$ in the vicinity of $R= 2 \mu$)
the existence of a (regular) horizon at  $R = 2 \mu$ for the metric (\ref{i.3}).
The second condition (\ref{i3.1b}) ensures  asymptotical (for $R \to
+\infty$) flatness of the metric.

Equations (\ref{i3.1}) may be rewritten in the following form:
\beq \label{i2.1}
  \frac{d}{dz} \left[
  \left(1 - z \right) \frac{d y^s}{dz} \right] =
          - K_s  q_s^2 \exp(- \sum_{l =1,2} A_{sl} y^l ),
\eeq
 $s = 1,2$. Here and in the following we use the following notations:
 $y^s= \ln H_s$,  $z = 2 \mu/R$, $q_s = Q_s/(2\mu)$ and  $K_s = h_s^{-1}$
 for $s = 1,2$, respectively.
We are seeking solutions to equations (\ref{i2.1}) for  $z \in (0,1)$
obeying
 \bearr \label{i2.1b}
   y^s(0) = 0, \yyy  \label{i2.1c}
   y^s(1) = y^{s}_0,
    \ear
where $y^{s}_0 = \ln H_{s0}$ are finite (real) numbers, $s = 1,2$.
Here $z=0$ (or, more precisely $z=+ 0$) corresponds to infinity ($R = + \infty$), while
$z=1$ (or, more rigorously, $z=1-0$ ) corresponds to the horizon ($R = 2 \mu$).

Equations (\ref{i2.1}) with  conditions of the finiteness on the horizon (\ref{i2.1c}) imposed
 imply the following  integral of motion:
  \bearr \label{i2.1d}
 \frac{1}{2}(1 -z)  \sum_{s,l =1,2} h_s A_{sl} \frac{d y^s}{dz} \frac{d y^l}{dz}
    +  \sum_{s =1,2} h_s \frac{d y^s}{dz}
              \nnn
           -  \sum_{s = 1,2} q_s^2 \exp(- \sum_{l =1,2} A_{sl} y^l)   = 0.
 \ear
Equations (\ref{i2.1}) and (\ref{i2.1c}) appear for  special
solutions to Toda-type equations  \cite{IMp2,IMp3,IMtop}
\beq \label{i2.1T}
  \frac{d^2 z^s}{du^2}   =    K_s Q_s^2 \exp(\sum_{l =1,2} A_{sl} z^l ),
 \eeq
for  functions
\beq \label{i2.z}
z^s(u) = - y^s -   \mu b^s u,
\eeq
$s = 1,2$, depending
on the harmonic radial variable $u$: $\exp(- 2 \mu u) = 1 -z$,
with  the following asymptotical
behavior for $u \to + \infty$ (on the horizon) imposed:
  \beq \label{i2.1as}
  z^s(u) = - \mu b^s u + z^s_{0} + o(1),
  \eeq
  where $z_{s0}$ are constants,  $s = 1,2$. Here and in the following we denote
  \beq \label{i2.b}
   b^s = 2 \sum_{l =1,2} A^{sl},
  \eeq
  where the inverse matrix $(A^{sl}) =  (A_{sl})^{-1}$ is well defined due
  to $\lambda_1 \neq - \lambda_2$.
  This follows from the relations
   \beq \label{i2.B}
      A_{sl} = 2 B_{sl} h_l, \qquad  B_{sl} = \frac{1}{2}
       + \varepsilon \chi_s \chi_l \lambda_s \lambda_l,
     \eeq
   where $\chi_1 = +1 $, $\chi_2 = - 1 $
   and the invertibility of the matrix $(B_{sl})$ for $\lambda_1 \neq - \lambda_2$,
   due to the relation $\det (B_{sl}) = \frac{1}{2} \varepsilon (\lambda_1 + \lambda_2)^2$.

   The  energy  integral of motion for (\ref{i2.1T}), which is compatible with the asymptotic
 conditions (\ref{i2.1as}),
 \bearr \label{i2.1ET}
   E = \frac{1}{4} \sum_{s,l =1,2} h_s A_{sl} \frac{d z^s}{du} \frac{d z^l}{du}          
    \yyy \nonumber       
     - \frac{1}{2} \sum_{s=1,2}  Q_s^2 \exp(\sum_{l =1,2} A_{sl} z^l) 
      =  \frac{1}{2} \mu^2 \sum_{s=1,2} h_s b^s, 
  \ear
   leads  to eq.  (\ref{i2.1d}).

 The derivation of the solutions (\ref{i.3})-(\ref{i.3bm}), (\ref{i3.1})-(\ref{i3.1b})
 may be extracted from the  relations of \cite{IMp1,IMp2,IMp3}, where the solutions with a horizon were obtained from 
 general spherically symmetric solutions governed by Toda-like  equations corresponding to a non-degenerate  (quasi-Cartan) matrix $A$. In our case 
 these equations are given by (\ref{i2.1T}) with the matrix $A$ from (\ref{i2.B}) and
 the condition   ${\rm det} A \neq 0$   implies $\lambda_1 \neq - \lambda_2$. The master equations
 (\ref{i3.1}) are equivalent to these Toda-like  equations.

% ====================================
\section{Integrable cases}

Explicit analytical solutions
to  eqs. (\ref{i3.1}), (\ref{i3.1a}), (\ref{i3.1b}) do not exist.
One may try to seek the solutions in the form

\beq \label{i3.12}
  H_{s} = 1 + \sum_{k = 1}^{\infty} P_s^{(k)}
  \left(\frac{1}{R}\right)^k,
\eeq
where $P_s^{(k)}$ are constants, $k = 1,2,\ldots, $ and
$s =1,2$, but only in few integrable cases the chain of
 equations for $P_s^{(k)}$ is dropped.

For $\varepsilon = + 1$,  there exist at least four integrable configurations
related to the Lie algebras $A_1 + A_1$, $A_2$,  $B_2 = C_2$ and $G_2$.

\subsection{$(A_1 + A_1)$-case}

Let us consider the case $\varepsilon = 1$ and
\beq \label{i4.0}
    \left(A_{ss'}\right)=
  \left( \begin{array}{*{6}{c}}
     2 &  0\\
     0 & 2\\
\end{array}
\right) .
\eeq
We obtain
\beq \label{i4.1}
  \lambda_1 \lambda_2 = \frac{1}{2}.
\eeq

For $\lambda_1 = \lambda_2$
we get a dilatonic coupling corresponding to string induced
model. The matrix (\ref{i4.0}) is the
Cartan matrix for the Lie algebra $A_1 + A_1$ ($A_1 = sl(2)$).
In this case
\beq \label{i4.2}
 H_s = 1 + \frac{P_s}{R},
\eeq
where
\beq \label{i4.3}
 P_s (P_s + 2 \mu) = K_s Q_s^2,
\eeq
$s = 1,2$. For   positive roots
of (\ref{i4.3})
\beq \label{i4.3p}
    P_s  = P_{s+} =  - \mu + \sqrt{\mu^2 + K_s Q^2_s},
\eeq
we are led to a well-defined solution for $R > 2\mu$  with asymptotically
flat metric  and  horizon at $R = 2 \mu$. We note that in the case
$\lambda_1 = \lambda_2$ the  $(A_1 + A_1)$-dyon
solution has a composite analog which was considered earlier in \cite{GM,ChHsuL};
see also \cite{Br0} for  certain generalizations.

\subsection{$A_2$-case}

Now we put $\varepsilon = 1$ and
\beq \label{i4.4a}
    \left(A_{ss'}\right)=
  \left( \begin{array}{*{6}{c}}
     2 &  -1\\
     -1 & 2\\
\end{array}
\right).
\eeq
We get
\beq \label{i4.4}
 \lambda_1 = \lambda_2 = \lambda, \qquad  \lambda^2 = 3/2. \qquad
\eeq
This value of dilatonic coupling constant appears after reduction to four dimensions of the 5d
Kaluza-Klein model. We get $h_s = 1/2$ and (\ref{i4.4a})
is the Cartan matrix for the Lie algebra $A_2 = sl(3)$.
In this case we obtain \cite{IMp2}

\beq \label{i4.5}
H_s = 1 + \frac{P_s}{R} + \frac{P_s^{(2)}}{R^2},
\eeq
where
\bearr \label{i4.6}
 2 Q_s^2 = \frac{P_s (P_s + 2 \mu) (P_s + 4 \mu)}{P_1 + P_2 + 4 \mu},
 \yyy \label{i4.6a}
 P_s^{(2)} = \frac{P_s (P_s + 2 \mu) P_{\bar{s}}}{2 (P_1 + P_2 + 4 \mu)},
\ear
$s = 1,2$; $\bar{s} = s+ 1 ({\rm mod} \ 2) = 2,1$.

In the composite case \cite{ABDI} the Kaluza-Klein uplift to $D=5$ gives us
the well-known Gibbons-Wiltshire solution  \cite{GibW},
which follows from the general spherically symmetric
dyon solution (related to $A_2$ Toda chain) from ref.  \cite{Lee}.

\subsection{$C_2$  case}

Now we put $\varepsilon = 1$ and
\beq \label{i4.7ac}
    \left(A_{ss'}\right)=
  \left( \begin{array}{*{6}{c}}
     2 &  -1\\
     -2 & 2\\
    \end{array} \right).   
\eeq
We  get integrable configuration, corresponding
to the Lie algebras $B_2 = C_2$ with the degrees of polynomials $(3,4)$. 
From (\ref{i.16a}),   (\ref{i.16b}) and (\ref{i4.7ac}) 
 we get the following relations for the dilatonic couplings:
\beq \label{i4.7aac}
 \frac{1}{2} +  \lambda_2^2 = 2 \left(\frac{1}{2} +  \lambda_1^2 \right),
 \quad  1 - 2 \lambda_1 \lambda_2  = - \frac{1}{2} -  \lambda_2^2.
 \eeq
 
Solving eqs. (\ref{i4.7aac}) we get 
$(\lambda_1, \lambda_2) = \pm (\sqrt{2}, \frac{3}{\sqrt{2}})$.

The moduli functions read \cite{GrIK}
\bearr \label{5.1v}
 H_1 = 1+P_1 z+P_1^{(2)}z^2+P_1^{(3)}z^3  \nnn
 = 1+ \bar{P}_1 \bar{z}+ \bar{P}_1^{(2)} \bar{z}^2+ \bar{P}_1^{(3)} \bar{z}^3, \yyy
 \label{5.2v}
H_2 = 1+ P_2 z+ P_2^{(2)}z^2+ P_2^{(3)}z^3+ P_2^{(4)} z^4  \nnn
     =  1+ \bar{P}_2 \bar{z} + \bar{P}_2^{(2)} \bar{z}^2+
        \bar{P}_2^{(3)} \bar{z}^3+ \bar{P}_2^{(4)} \bar{z}^4,   
 \ear
where $P_s= P_s^{(1)} = \bar{P}_s (2 \mu)$ and $P_s^{(k)} = \bar{P}_s^{(k)} (2 \mu)^k$ are constants, 
$s = 1,2$, and $z = 1/R$; $\bar{z} = 2 \mu / R$.

For parameters $\bar{B}_s = - K_s Q_s^2/(2 \mu)^2$  we get the following relations \cite{GrIK}
 \bearr \label{5.3v}
 2 \bar{B}_1 = -\Delta + (2 \bar{P}_1+ 3)(2 + \bar{P}_2),
 \yyy
  \label{5.4v}
    \bar{B}_2 = \Delta -  2 - 2 \bar{P}_1 (\bar{P}_1 + 3)-(2+ \bar{P}_2)^2,
\ear
and for parameters $\bar{P}_s^{(k)}$ we obtain \cite{GrIK}
\bearr \label{5.5v}
 4 \bar{P}_1^{(2)}= 6 + 3 \bar{P}_2 - \Delta + 2 \bar{P}_1 (3 + \bar{P}_1 + \bar{P}_2), 
 \yyy  
12 \bar{P}_1^{(3)}= - \Delta (2 + \bar{P}_1 + \bar{P}_2)+ 12 
                     +18 \bar{P}_1  
 \nnn     + 2 \bar{P}_1^3 + 3 \bar{P}_2(4 + \bar{P}_2)
        \label{5.6v}
\yyy
        + 2 \bar{P}_1^2(5 + \bar{P}_2) + \bar{P}_1 \bar{P}_2 (11 + 2 \bar{P}_2),
\nnn
  \label{5.7v}
  2 \bar{P}_2^{(2)}= -6- 2 \bar{P}_1(3 + \bar{P}_1)- 3 \bar{P}_2 + \Delta, 
 \yyy  
    6 \bar{P}_2^{(3)}= \Delta (2 + 2 \bar{P}_1+ \bar{P}_2)- 12 
 \nnn   
    -  24 \bar{P}_1  - 4 \bar{P}_1^3 
    - 3 \bar{P}_2(4 + \bar{P}_2)
 \nnn  
    - 2 \bar{P}_1 \bar{P}_2(7 + \bar{P}_2)
     - 2 \bar{P}_1^2(8 + \bar{P}_2), \label{5.8v}
 \yyy 
    \label{5.9v}
   24 \bar{P}_2^{(4)}= \Delta [2 \bar{P}_1^2+ (3 + \bar{P}_2)(2 + 2 \bar{P}_1+ \bar{P}_2)]
   -4 \bar{P}_1^4
 \nnn    
     - 3(2 + \bar{P}_2)^2(3 + \bar{P}_2)- 2 \bar{P}_1(3+ \bar{P}_2)^2(4 + \bar{P}_2)
 \nnn
   -  4 \bar{P}_1^3(6 + \bar{P}_2)- \bar{P}_1^2(60 + 30 \bar{P}_2+4 \bar{P}_2^2),
\ear
where
 \beq \label{5.10v}
\Delta = \sqrt{4\left(3+ \bar{P}_1(3+\bar{P}_1)\right)^2+(3+2 \bar{P}_1)^2 \bar{P}_2(4+ \bar{P}_2)}.
\eeq
It may be verified that $\bar{B}_1 < 0$ and $\bar{B}_2 < 0$ for $\bar{P}_1 > 0$, $\bar{P}_2 > 0$.

\subsection{$G_2$ case}

If we put $\varepsilon = 1$ and
\beq \label{i4.7ag}
    \left(A_{ss'}\right)=
  \left( \begin{array}{*{6}{c}}
     2 &  -1\\
     -3 & 2\\
    \end{array} \right),
\eeq
we also get integrable configuration, corresponding
to the Lie algebra $G_2$, respectively, with the degrees of polynomials $(n_1,n_2) =(6,10)$. 
From (\ref{i.16a}),   (\ref{i.16b}) and (\ref{i4.7ag}) 
 we get the following relations for the dilatonic couplings:
\beq \label{i4.7aag}
 \frac{1}{2} +  \lambda_2^2 = 3 \left(\frac{1}{2} +  \lambda_1^2 \right),
 \  1 - 2 \lambda_1 \lambda_2  = - \frac{1}{2} -  \lambda_2^2.
 \eeq
 
Solving eqs. (\ref{i4.7aag}) we get 
$(\lambda_1, \lambda_2) = \pm \left(\frac{5}{\sqrt{6}}, 3\sqrt{\frac{3}{2}}\right)$.

Due to ref. \cite{I-14} our polynomials $H_1$ and $H_2$ may be calculated
by using so-called fluxbrane polynomials which  obey the equations
  \beq \label{4.1v}
  \frac{d}{dz} \left( \frac{z}{{\cal H}_s} \frac{d}{dz} {\cal H}_s \right) =
   n_s p_s  \prod_{l = 1}^{2}  {\cal H}_{l}^{- A_{s l}},
  \eeq
 with  the  boundary conditions imposed
 \beq \label{4.2v}
   {\cal H}_{s}(+ 0) = 1.
\eeq  

For $G_2$-case these polynomials read \cite{GonIM}
 
\bearr \label{G.2v}
 {\cal H}_{1} = 1+ 6 p_1 z+ 15 p_1 p_2 z^2 + 20 p_1^2 p_2 z^3 + \nnn
   \qquad  15 p_1^3 p_2 z^4 + 6 p_1^3 p_2^2 z^5 + p_1^4 p_2^2 z^6 , \yyy
 \label{G.3}
  {\cal H}_2 =  1+  10 p_2 z + 45 p_1 p_2 z^2  +  120 p_1^2 p_2 z^3
  \nnn   \qquad +  p_1^2 p_2 (135 p_1 + 75 p_2) z^4  + 252 p_1^3 p_2^2 z^5
  \nnn
   + p_1^3 p_2^2 \biggl(75 p_1 + 135 p_2 \biggr)z^6   +  120 p_1^4 p_2^3  z^7
   \nnn
   + 45 p_1^5 p_2^3 z^8 +  10 p_1^6 p_2^3 z^9
      + p_1^{6} p_2^{4}  z^{10}.
  \ear
 
Let us denote $f = f(z)= 1 - 2\mu z$, $z = 1/R$.
Then the relations (\ref{i3.1}) may be rewritten as 
\beq \label{5.1vv}
 \frac{d}{df} \left( \frac{f}{H_s}
 \frac{d}{df} H_s \right) =  B_s (2 \mu )^{-2}
 \prod_{l =1}^{2}  H_{l}^{- A_{s l}}, 
 \eeq
  $B_s  =  - K_{s} Q_s^2$, $s = 1,2$.
 These relations could be solved by using fluxbrane
polynomials ${\cal H}_{s}(f) = {\cal H}_{s}(f; p)$, corresponding
to $2 \times 2$ Cartan matrix $(A_{s l})$, where $p =
(p_1,p_2)$ is the set of  parameters. Here
we impose the restrictions $p_s \neq 0$ for all $s$.

Due to approach of ref. \cite{I-14} (see also \cite{GalZad}) we put 
\beq \label{5.2vv}
  H_s(z) = {\cal H}_{s}(f(z);p)/{\cal H}_{s}(1;p)
\eeq 
for  $s = 1,2$. Then the relations (\ref{5.1vv}), or, equivalently, (\ref{i3.1})
are satisfied identically if \cite{I-14}

\beq \label{5.3vv}
 n_s p_s  \prod_{l =1}^{2}  ({\cal H}_{l}(1;p))^{- A_{s l}} = B_s /(2 \mu)^2,
 \eeq 
$s = 1,2$; where $n_1 = 6$ and $n_2 = 10$.

We call the set of parameters $p = (p_1,p_2)$ ($p_i \neq 0$)  as proper one if \cite{I-14}
  \beq \label{5.4vv}
 {\cal H}_{s}(f;p) > 0
 \eeq
 for all $f \in [0,1]$ and $s = 1,2$.  In what  follows we consider only proper $p$.
 Relations (\ref{5.3vv})  $p_s < 0$ and $B_s < 0$ for $s = 1,2$.

The  boundary conditions  (\ref{i3.1a}) are valid since
  \beq \label{5.6vv}
  H_{s}((2\mu)^{-1} -0) = 1/{\cal H}_{s}(1;p) > 0,
  \eeq
 $s = 1,2$,  and conditions (\ref{i3.1b}) are satisfied just due to
 definition (\ref{5.2v}).

 Locally, for small enough
 $p_i$ the relation  (\ref{5.3vv}) defines one-to-one correspondence between
 the sets of parameters $(p_1,p_2)$ and $(Q_1^2, Q_2^2)$ and the 
 set $(p_1,p_2)$ is proper.

\subsection{Special solution with dependent charges}

There exists also a special solution
\beq \label{i4.7}
 H_s = \left(1 + \frac{P}{R}\right)^{b^s},
\eeq
with $P > 0$ obeying
\beq \label{i4.8a}
  \frac{K_s}{b_s} Q_s^2 = P (P + 2 \mu),
\eeq
 $s = 1,2$.
 Here  $b^s \neq 0$ is defined in (\ref{i2.b}).
 This solution is a special case of more general
 ``block orthogonal'' black brane  solutions  \cite{Br,IMJ2,CIM}.

 The calculations give us the following relations:
 \beq \label{i4.8b}
  b^s = \frac{2 \lambda_{\bar s}}{\lambda_1 + \lambda_2} K_s,
 \eeq

 \beq \label{i4.8c}
    Q_s^2 \frac{(\lambda_1 + \lambda_2)}{2 \lambda_{\bar s}}
    = P (P + 2 \mu) = \frac{1}{2} Q^2,
 \eeq
 where $s = 1, 2$ and ${\bar s} = 2,1$, respectively.
 Our solution is well defined if $\lambda_1 \lambda_2 > 0$,
 i.e. the two  coupling constants have the same sign.

 For positive roots of (\ref{i4.8c})
\beq \label{i4.8p}
    P  = P_{+} =  - \mu + \sqrt{\mu^2 + \frac{1}{2}Q^2}
\eeq
we get for $R > 2\mu$ a well-defined  solution with asymptotically
flat metric  and   horizon at $R = 2\mu$ which is valid for
both signs $\varepsilon = \pm 1$.

By changing the radial variable, $r = R + P$, we get \cite{ABI}
 \bearr \label{i4.12}
  ds^2 =   - f(r) dt^2 +   f(r)^{-1} dr^2 + r^2  d \Omega^2_{2},
     \yyy    \label{i4.12F}
  F^{(1)}= \frac{Q_1}{r^2} dt \wedge dr, \quad F^{(2)}  =  Q_2 \tau,
  \quad \varphi = 0,
    \ear
 where $f(r) = 1 - \frac{2GM}{r} + \frac{Q^2}{2r^2}$, $ Q^2 = Q_1^2 + Q_2^2$ and $GM = P + \mu$ =
 $\sqrt{\mu^2 + \frac{1}{2} Q^2} > \frac{1}{\sqrt{2}} |Q|$ and
 
 \beq  \label{i4.11}
   Q_1^2 = \frac{\lambda_{2}}{\lambda_1 + \lambda_2} Q^2, \qquad
   Q_2^2 = \frac{ \lambda_{1}}{\lambda_1 + \lambda_2} Q^2.
 \eeq

 The metric in these variables is coinciding with the well-known Reissner-Nordstr\"om  
 metric governed by two  parameters: $GM > 0$ and $ Q^2 < 2 (GM)^2$. 
 We have two horizons in this case.  Electric and magnetic charges are not independent but obey eqs.
 (\ref{i4.11}).
 
 {\bf $H_2(q,q)$ case.} It should be noted that for the case  
 \beq \label{i4.13}
    \lambda_1 = \lambda_2 = \lambda, \quad \lambda^2 = \frac{q + 2}{2(q - 2)},
     \quad \varepsilon = -1, 
  \eeq
 $q = 2,3, 4, \dots$, we get a special solution with $Q_1^2 = Q_2^2 = Q^2/2$ and 
 \beq \label{i4.14}
     \left(A_{ss'}\right)=
   \left( \begin{array}{*{6}{c}}
       2   &  -q \\
      -q   &   2 \\
 \end{array}
 \right), 
 \eeq
which is the Cartan matrix of the hyperbolic Kac-Moody algebra $H_2(q,q)$ ($q = 3,4,5, \dots $), see 
\cite{I-Sym-17} and references therein.  

\subsection{The limiting $A_1$-cases}

In the following we will use two limiting solutions: an electric one
with $Q_1 = Q \neq 0$ and   $Q_2 = 0$,
\beq \label{i4.e}
H_1 = 1 + \frac{P_1}{R}, \qquad H_2 = 1,
\eeq
and a magnetic one  with $Q_1 = 0$ and $Q_2 = Q \neq 0$,
 \beq \label{i4.m}
H_1 = 1, \qquad  H_2 = 1 + \frac{P_2}{R}.
 \eeq
In both cases $P_s  =   - \mu + \sqrt{\mu^2 + K_s Q^2}$.
These solutions correspond to  the Lie algebra $A_1$.
In various notations the solution  (\ref{i4.e}) appeared earlier in \cite{BronShikin,Hein,GM}, 
and it was extended to the multidimensional
case in  \cite{Hein,GM,BBFM,BI}. The special case with $\lambda^2_1 = 1/2$, $\varepsilon= 1$,
was considered earlier in \cite{Gibbons,GHS}.

% ====================================
\section{Physical parameters}

Here we consider certain physical parameters
corresponding to the solutions under consideration.

\subsection{ADM mass and scalar charge}

For ADM gravitational mass we get from (\ref{i.3})
\beq \label{i5.1}
 GM =   \mu +  \frac{1}{2} (h_1 P_1 + h_2 P_2),
\eeq
where  the parameters $P_s = P_s^{(1)}$ appear in
eq. (\ref{i3.12}) and $G$ is the gravitational constant.

The scalar charge just follows  from (\ref{i.3a}):
\beq \label{i5.1s}
 Q_{\varphi} =  \varepsilon (\lambda_1 h_1 P_1 -  \lambda_2 h_2 P_2).
\eeq

For the special solution (\ref{i4.7})  with $P >0$ we get

\beq \label{i5.1sim}
  GM =  \mu + P = \sqrt{\mu^2 + Q^2}, \quad
  Q_{\varphi} = 0.
 \eeq

For fixed charges $Q_s$ and the extremality parameter $\mu$
the mass $M$ and scalar charge $Q_{\varphi}$ are not
independent but obey a certain constraint. Indeed,
for fixed parameters $P_s = P_s^{(1)}$ in   (\ref{i3.12}) we
get
\beq \label{i5.12}
      y^s = \ln H_s = \frac{P_s}{2\mu} z + O(z^2),
 \eeq
for   $z \to + 0$, which after substitution into  (\ref{i2.1d}) gives  (for $z =0$)
the following identity:

\beq \label{i5.1p}
 \frac{1}{2} \sum_{s,l =1,2} h_s A_{sl} P_s P_l
 + 2 \mu \sum_{s =1,2} h_s P_s  =  \sum_{s =1,2} Q_s^2.
\eeq

By using  eqs. (\ref{i5.1}) and (\ref{i5.1s}) this identity may be rewritten
in the following form:

\beq \label{i5.1id}
     2 (GM)^2   +   \varepsilon  Q_{\varphi}^2   = Q_1^2 + Q_2^2 + 2 \mu^2.
\eeq

It is remarkable that this formula does not contain $\lambda$.
We note that in the extremal case $\mu = +0$ this relation for $\varepsilon = 1$
 was obtained earlier  in \cite{PTW}.

\subsection{Hawking temperature and  entropy}

The Hawking temperature corresponding to
the solution is found to be
  \beq \label{i5.2}
 T_H=   \frac{1}{8 \pi \mu}  H_{10}^{- h_1} H_{20}^{- h_2},
 \eeq
where $H_{s0}$ are defined in (\ref{i3.1a}).
Here and in the following we put $c= \hbar = \kappa =1$.

For special solutions (\ref{i4.7})   with $P >0$ we get
\beq \label{i5.3sim}
  T_H =  \frac{1}{8 \pi \mu} \left(1 + \frac{P}{2 \mu}\right)^{-2}.
 \eeq
In this case the Hawking temperature $T_H$ does not depend upon
 $\lambda_s$ and  $\varepsilon$, when $\mu$ and $P$ (or $Q^2$) are fixed.

The Bekenstein-Hawking (area) entropy $S = A/(4G)$,
corresponding to the horizon at $R = 2\mu$, where $A$ is the horizon area,  reads
\beq \label{i5.2s}
S_{BH} =   \frac{4 \pi \mu^2}{G}  H_{10}^{h_1} H_{20}^{h_2}.
\eeq
It follows from (\ref{i5.2}) and (\ref{i5.2s}) that the product

\beq \label{i5.2st}
  T_H  S_{BH} =   \frac{\mu}{2G}
\eeq
does not depend upon  $\lambda_s$,  $\varepsilon$ and the charges $Q_s$.
This product does not use an explicit form of the moduli functions $H_s(R)$.

Using (\ref{i5.1id}) and (\ref{i5.2st}) we get a sort of Smarr relation
\beq \label{i5.1smarr}
  2 (GM)^2   +   \varepsilon  Q_{\varphi}^2   = Q_1^2 + Q_2^2 + 8 (G T_H  S_{BH} )^2.
\eeq

\subsection{PPN parameters}

Introducing a new radial variable $\rho$ by the relation
 $R =   \rho (1 + (\mu/2\rho))^2$
($\rho > \mu/2$),  we obtain  the 3-dimensionally
conformally flat form of the metric  (\ref{i.3})
\bearr \label{i5.3}
g =   U \Biggl\{ -
U_1 \frac{\left(1 - (\mu/2\rho) \right)^2}
{\left(1 + (\mu/2\rho) \right)^2} dt \otimes dt 
  \nnn
  \qquad   + \left(1 + \frac{\mu}{2 \rho} \right)^4
         \delta_{ij} dx^i \otimes dx^j \Biggr\},
\ear
where  $\rho^2 =  |x|^2 =   \delta_{ij}x^i x^j$ ($i,j =   1,2,3$)
and
\beq \label{5.5.1a}
U =   \prod_{s = 1,2} H_s^{h_s},\qquad U_1 =   \prod_{s  = 1,2} H_s^{-2 h_s}.
\eeq

The parametrized post-Newtonian (PPN) parameters
$\beta$ and $\gamma$ are defined by the following standard relations:
\bearr \label{A.1}
g_{00} =   - (1 -  2 V + 2 \beta V^2 ) + O(V^3),
\yyy
\label{A.2}
g_{ij} =   \delta_{ij}(1 + 2 \gamma V ) + O(V^2),
\ear
$i,j =   1,2,3$, where $V =   GM/\rho$
is  Newton's potential, $G$ is the gravitational constant and
$M$ is the gravitational mass (for our case see (\ref{i5.1})).

The calculations of PPN (or Eddington) parameters for the metric (\ref{i5.3})
give us \cite{ABI}:
\beq \label{i5.8}
\beta  = 1 +   \frac{1}{4(GM)^2}  (Q_1^{2} + Q_2^{2}), \qquad \gamma = 1.
\eeq

These parameters  do not depend upon $\lambda_s$
and $\varepsilon$. They may be calculated just without knowledge of the explicit
relations for  the moduli functions $H_s(R)$.

 These parameters (at least  formally)
obey  the observational restrictions for the solar system \cite{Wil},
when  $Q_s/(2GM)$ are small enough.

% ====================================
\section{Bounds on  mass and scalar charge}

Here we outline the following hypothesis, which is supported by certain numerical calculations \cite{ABDI}.
For $h_1 = h_2$ this conjecture was proposed in ref. \cite{ABDI}.

{\bf Conjecture.}
{\em For any $h_1 >0$, $h_2 >0$, $\varepsilon = \pm 1$, $Q_1 \neq 0$, $Q_2\neq 0$ and $\mu > 0$:
(A) the moduli functions $H_s(R)$, which obey  (\ref{i3.1}), (\ref{i3.1a}) and (\ref{i3.1b}),
are uniquely defined and hence the parameters $P_1$, $P_2$, the gravitational mass $M$ and
the scalar charge $Q_{\varphi}$ are uniquely defined too;
(B) the parameters $P_1$, $P_2$ are positive and the  functions $P_1 = P_1(Q_1^2,Q_2^2)$,
 $P_2 = P_2(Q_1^2,Q_2^2)$  define a diffeomorphism of $\R_{+}^2$ ($\R_{+} = \{x| x>0 \}$);
  (C) in the limiting case we have: (i) for $Q_2^2 \to + 0$:
  $P_1 \to   - \mu + \sqrt{\mu^2 + K_1 Q_1^2}$, $P_2 \to +0$
  and (ii) for $Q_1^2 \to + 0$: $P_1 \to +0$, $P_2 \to   - \mu + \sqrt{\mu^2 + K_2 Q_2^2}$.}

 The  conjecture could be readily verified  for the $(A_1 + A_1)$-case  $\varepsilon = 1$,
$\lambda_1 \lambda_2 = 1/2$.
 Another integrable $A_2$-case  $\varepsilon = 1$, $\lambda_1 = \lambda_2 = \lambda$, 
 $\lambda^2 = 3/2$ is more involved.

Let us define 
$h_{min} = {\rm min} (h_1,h_2)$, $h_{max} = {\rm max} (h_1,h_2)$, and
  $|\lambda|_{max} = { \rm max } (|\lambda|_{1}, |\lambda|_{2} )$; then we get
  $h_{min} = (\frac{1}{2} +   |\lambda|^2_{max})^{-1}$ for $\varepsilon = +1$ and
  $h_{max} = (\frac{1}{2} -   |\lambda|^2_{max})^{-1}$ for $\varepsilon = -1$.

The Conjecture implies the following proposition.

{\bf Proposition 2 \cite{ABI}.}
{\em In the framework of the conditions of Proposition 1, 
 the following bounds on the mass and scalar charge are  valid for all
$\mu >0$:
\bearr \label{i5.16p}
   \frac{1}{2}\sqrt{h_{min} (Q_1^2 + Q_2^2)}  < GM, \yyy \label{i5.16sp}
    |Q_{\varphi}| < |\lambda|_{max} \sqrt{h_{min} (Q_1^2 + Q_2^2)},
  \ear
for $\varepsilon = +1$ $(0< h_s < 2)$, and
\bearr \label{i5.16m}
 \sqrt{\frac{1}{2} (Q_1^2 + Q_2^2)} < GM,
 \yyy \label{i5.16sm}
  |Q_{\varphi}| < |\lambda|_{max} \sqrt{h_{max} (Q_1^2 + Q_2^2)},
  \ear
for $\varepsilon = -1$ $(h_s > 2)$}.

In ref. \cite{ABDI} Proposition  was  proved  for the case $\lambda_1 = \lambda_2$ ($h_1 = h_2$).
In this case the bound (\ref{i5.16p}) is coinciding (up to notations) with  the bound (6.16) from ref. \cite{GKLTT} (BPS-like inequality), which was proved there  by using certain  spinor techniques.

We note that here we were dealing with a  special class of solutions with phantom scalar field ($\varepsilon = -1$). Even in the  limiting case  $Q_2 = +0$ and $Q_1 \neq 0$ there exist  phantom black hole solutions which are not covered by our analysis, see refs. \cite{CFR,ACFR}.

 When one of  $h_s$, say $h_1$, is negative, 
the  Conjecture is not valid. This may be verified just by analyzing the solutions with small enough charge $Q_2$.

% ====================================
\section{Conclusions}

In this paper a family of non-extremal black hole dyon-like
 solutions in a 4d gravitational model
with a scalar field and two Abelian vector fields
 is overviewed. The scalar field is
either ordinary ($\varepsilon = +1$) or phantom  ($\varepsilon = -1$).
The model contains two dilatonic coupling constants
$\lambda_s \neq 0$, $s =1,2$, obeying $\lambda_1 \neq - \lambda_2$.

The solutions are defined up to two moduli functions $H_1(R)$ and
$H_2(R)$,  which obey two differential equations  of  second order
with boundary conditions imposed. For $\varepsilon = +1$ these
equations are integrable for four cases, corresponding
to the Lie algebras $A_1 + A_1$, $A_2$, $B_2 = C_2$ and $G_2$. 
The solutions are presented here.

There is also a special subclass of solutions with dependent
electric and magnetic charges: $\lambda_1 Q_1^2 = \lambda_2 Q_2^2$, 
which is defined for all (admissible)  $\lambda_s$ and 
$\varepsilon$ obeying $\lambda_1 \lambda_2 > 0$. It is shown that this 
subclass contains  solutions corresponding to hyperbolic Kac-Moody algebras 
$H_2(q,q)$, $q = 3,4, \dots$.

Here we have also derived some physical parameters of the solutions:
gravitational mass $M$, scalar charge $Q_{\varphi}$, Hawking temperature,
black hole area entropy and post-Newtonian parameters  $\beta$,  $\gamma$.
The PPN parameters $\gamma =1$ and $\beta$ do not depend upon
$\lambda_s$ and $\varepsilon$, if the values of $M$ and $Q_{\varphi}$ are fixed.

We have also considered  a formula, which relates  $M$,  $Q_{\varphi}$,
the dyon charges $Q_1$, $Q_2$, and the extremality parameter $\mu$
 for all values of  $\lambda_s \neq 0$. Remarkably, this formula does not contain $\lambda_s$
 and coincides with that of ref. \cite{ABDI}.
As in the case $\lambda_1 = \lambda_2$, the product of the Hawking temperature and
the Bekenstein-Hawking entropy do not depend upon $\varepsilon$,  $\lambda_s$
and  the moduli functions $H_s(R)$.

 Here we have presented  lower bounds on the gravitational mass and upper bounds on  the scalar charge for  $1 +2 \lambda^2_s \varepsilon > 0$,  which are based on the conjecture 
 on the parameters of solutions $P_1 = P_1(Q_1^2,Q_2^2)$,  $P_2 = P_2(Q_1^2,Q_2^2)$.
  For $\lambda_1 = \lambda_2$ the conjecture is supported by results of numerical calculations from ref. \cite{ABDI}. 
 A rigorous proof of this  conjecture may be a subject of a separate publication.
 For $\varepsilon  = +1$ and $\lambda_1 = \lambda_2$  the lower bound on the gravitational  mass 
 is in agreement for  with that obtained earlier by Gibbons et al. in ref. \cite{GKLTT} by using certain  spinor techniques. 
 
We note that there exist conditions on the dilatonic coupling constants $\lambda_s$
which guarantee the existence of the second (hidden) horizon and the existence of the extremal black hole
in the  limit $\mu = +0$, see \cite{Dav,GalZad}. For  $\varepsilon  = +1$, $\lambda_1 = \lambda_2$ this problem
was analyzed in refs. \cite{PTW,GKO}. 

% -------------------------------------------

%\small
\subsection*{Funding}

The author Abishev M.E. acknowledge the support from the Program of target financing of the Ministry of Education and Science of the Republic of Kazakhstan, Grant No BR05236277. The author Toktarbay S. acknowledge the support from the Project Grant No AP05133630 of the Ministry of Education and Science of the Republic of Kazakhstan. 
The author Ivashchuk V. D. acknowledge the support from   the  Russian Foundation for Basic Research,  Grant  Nr. 19-02-00346.

} % end au-def
% =================================================================


\small

\begin{thebibliography}{99}


\bibitem{BronShikin}
K. A. Bronnikov and G. N. Shikin, ``On interacting fields in general relativity theory,''
Izvest. Vuzov (Fizika), {\bf 9}, 25-30 (1977) [in Russian];
Russ. Phys. J. {\bf 20}, 1138-1143 (1977).

\bibitem{Gibbons}
G. W. Gibbons, ``Antigravitating black hole solutions with scalar hair in $N=4$ supergravity,''
Nucl. Phys.  B {\bf 207}, 337-349 (1982).

\bibitem{Lee}
S.-C. Lee, ``Kaluza-Klein dyons and the Toda lattice,''  
Phys. Lett. B {\bf 149} (1-3), 98-99 (1984).

\bibitem{GibW}
G. W. Gibbons and D. L. Wiltshire, ``Spacetime as a membrane in
higher dimensions,'' Nucl. Phys. B {\bf 287}, 717-742 (1987).

\bibitem{Hein}
O. Heinrich, ``Charged black holes in compactified higher-dimensional Einstein-Maxwell
theory,''  Astron. Nachr. {\bf 309} (4), 249-251 (1988).

\bibitem{GM}
G. W. Gibbons and K. Maeda, ``Black holes and membranes in
higher-dimensional theories with dilaton fields,''  Nucl. Phys.
B {\bf 298}, 741-775 (1988).

\bibitem{GHS}
D. Garfinkle, G. Horowitz and A. Strominger, ``Charged black holes in string theory,''
Phys. Rev. D {\bf 43},  3140 (1991); Erratum ibid. {\bf 45}, 3888 (1992).

\bibitem{ChHsuL}
G.-J. Cheng, R. R. Hsu and W.-F. Lin, ``Dyonic Black Holes in String Theory,''
 J. Math. Phys. {\bf 35}, 4839-4847 (1994);  arXiv: hep-th/9302065.

\bibitem{GKLTT}
G. W. Gibbons, D. Kastor, L. A. J. London, P. K. Townsend and J. Traschen,
``Supersymmetric Self-Gravitating Solitons,''
Nucl. Phys. B  {\bf 416}, 850-880 (1994); arXiv: hep-th/9310118.


\bibitem{BBFM}
 U. Bleyer, K. A. Bronnikov, S. B. Fadeev and V. N. Melnikov,
 ``Black  hole stability in multidimensional gravity theory,''
 Astron. Nachr. {\bf 315} (4), 399-408 (1994);  arXiv: gr-qc/9405021.

 \bibitem{BI}
 U. Bleyer and V. D. Ivashchuk,
 ``Mass bounds for Multidimensional  Charged Dilatonic Black Holes,''
 Phys. Lett. B {\bf 332}, 292-296 (1994); arXiv: gr-qc/9405018.

\bibitem{PTW}
S. J. Poletti, J. Twamley and D. L. Wiltshire,
``Dyonic dilaton black holes,'' Class. Quant. Grav. {\bf 12}, 1753-1770 (1995); Erratum
ibid. {\bf 12}, 2355 (1995); arXiv: hep-th/9502054.

\bibitem{Br0}
K.A. Bronnikov,
``On spherically symmetric solutions in D-dimensional dilaton gravity,''
Grav. Cosmol. {\bf 1},  67-78 (1995); arXiv: gr-qc/9505020.

\bibitem{LP}
H. L\"u and C. N. Pope, ``p-brane Solitons in Maximal Supergravities,''
Nucl. Phys. B {\bf  465}, 127-156 (1996); arXiv: hep-th/9512012.

\bibitem{DLP}
M. J. Duff, H. Lu and C. N. Pope, ``The Black Branes of M-theory,''
Phys. Lett. B {\bf  382}, 73-80 (1996); arXiv: hep-th/9604052.

\bibitem{LPX}
H. L\"u, C. N. Pope  and K. W. Xu, ``Liouville and Toda solitons in
M-Theory,''  Mod. Phys. Lett. A {\bf  11}, 1785-1796 (1996);
arXiv: hep-th/9604058.

\bibitem{IMp1}
V. D. Ivashchuk and V. N. Melnikov, ``P-brane black holes
for general intersections,'' Grav.  Cosmol.
{\bf 5} (4), 313-318 (1999); arXiv: gr-qc/0002085.

\bibitem{IMp2}
V. D. Ivashchuk and V. N. Melnikov, ``Black hole p-brane solutions for
general intersection rules,''  Grav.  Cosmol.
{\bf 6} (1), 27-40 (2000); arXiv: hep-th/9910041.

\bibitem{IMp3}
V. D. Ivashchuk and V. N. Melnikov, ``Toda  p-brane black holes
and polynomials related to Lie algebras,'' 
Class. Quantum Grav. {\bf 17}, 2073-2092 (2000); arXiv: math-ph/0002048.

\bibitem{IMtop}
V. D. Ivashchuk and V. N. Melnikov, ``Exact solutions in
multidimensional gravity with antisymmetric forms'', topical review,
Class. Quantum Grav.  {\bf 18}, R1-R66 (2001);
arXiv: hep-th/0110274.

\bibitem{FIMS}
S. B. Fadeev, V. D. Ivashchuk, V. N. Melnikov and L. G. Sinanyan,
``On PPN parameters for dyonic black hole solutions,''
 Grav. Cosmol. {\bf 7} (4), 343-344 (2001).

\bibitem{CGLO}
G. Clement, D. Gal'tsov, C. Leygnac and D. Orlov,
``Dyonic branes and linear dilaton background,''
Phys. Rev. D {\bf 73}, 045018 (2006);
arXiv: hep-th/0512013.

\bibitem{GO}
D.V. Gal'tsov and D.G. Orlov,  
``Liouville and Toda dyonic branes: regularity and BPS limit,''
{\it Grav. Cosmol.} {\bf 11}: 235-243 (2005); arXiv: hep-th/0512345.

\bibitem{LY}
H. L\"u and W. Yang, ``SL(n,R)-Toda Black Holes,'' arXiv:
1307.2305.

\bibitem{Ifbb}
V. D. Ivashchuk, ``Black brane solutions governed by fluxbrane
polynomials,'' J.  Geom. Phys., {\bf 86}, 101-111
(2014); arXiv: 1401.0215.

\bibitem{GKO}
 D. Gal'tsov, M. Khramtsov and D. Orlov,
 ``Triangular'' extremal dilatonic dyons,''
 Phys. Lett. B {\bf 743}, 87-92 (2015);
 arXiv: 1412.7709.

\bibitem{ABDI}
M. E. Abishev, K. A. Boshkayev, V. D. Dzhunushaliev and V. D. Ivashchuk,
``Dilatonic dyon black hole solutions,''
Class. Quantum  Grav. {\bf 32} (16), 165010 (2015); arXiv: 1504.07657.

\bibitem{ABI}
  M. E. Abishev, K. A. Boshkayev, and V. D. Ivashchuk, 
  ``Dilatonic dyon-like black hole solutions in the model with two Abelian gauge fields,'' 
  Eur. Phys. J.  C  {\bf 77}, 180 (10 pages) (2017). 

\bibitem{Dav}
E. A. Davydov, ``Discreteness of dyonic dilaton black holes,'' arXiv: 1711.04198. 

\bibitem{GalZad}
A. Zadora, D. V. Gal'tsov, C.-M. Chen,
``Higher-$n$ triangular dilatonic black holes,''
Phys. Lett. B {\bf 779}, 249-256  (2018). 

\bibitem{KKLP}
N. Khviengia, Z. Khviengia, H. L\"u and C. N. Pope, 
``Towards a Field Theory of F-theory,''
Class. Quant. Grav. {\bf 15}, 759-773 (1998); 
arXiv: hep-th/9703012.

\bibitem{ArkH}
N. Arkani-Hamed, H.-Ch. Cheng, M. A. Luty and S. Mukoyama,
``Ghost Condensation and a Consistent Infrared Modification of Gravity,''
 JHEP {\bf 0405}, 074 (2004); arXiv: hep-th/0312099.

\bibitem{Kom}
E. Komatsu et al.,
``Seven-Year Wilkinson Microwave Anisotropy Probe (WMAP)
Observations: Cosmological Interpretation,''
Astrophys. J. Suppl.  {\bf 192}, 18 (2011);
arXiv: 1001.4538[astro-ph].

\bibitem{GrIK}
M. A. Grebeniuk, V. D. Ivashchuk and S.-W. Kim,
``Black-brane solutions for $C_2$ algebra,''
J. Math. Phys. {\bf 43}, 6016-6023 (2002);
arXiv: hep-th/0111219.

\bibitem{I-14}
V. D. Ivashchuk, ``Black brane solutions governed by fluxbrane
polynomials,''  J.  Geom. Phys., {\bf 86}, 101-111 (2014).

 \bibitem{GonIM} 
 I. S. Goncharenko, V. D. Ivashchuk and V. N. Melnikov,
  ``Fluxbrane and S-brane solutions with polynomials related to rank-2 Lie algebras'',
  Grav.  Cosmol. {\bf 13}  (4), 262-266 (2007);
  math-ph/0612079.

\bibitem{Br}
K. A. Bronnikov, ``Block-orthogonal Brane systems, Black
Holes and Wormholes,''  Grav.  Cosmol., {\bf 4} (1),  49
(1998); arXiv: hep-th/9710207.

\bibitem{IMJ2}
V.D. Ivashchuk and V.N. Melnikov,
Multidimensional Cosmological and Spherically Symmetric Solutions
with Intersecting $p$-branes,''  \\
In: Lecture Notes in Physics, v. 537. Mathematical and Quantum Aspects of
Relativity and Cosmology. Eds.: S. Cotsakis and G. Gibbons. Springer, Berlin,
2000, p. 214; arXiv: gr-qc/9901001.

\bibitem{CIM}
S. Cotsakis, V. D. Ivashchuk and V. N. Melnikov,
P-branes Black Holes and Post-Newtonian Approximation,''
 Grav. Cosmol. {\bf 5}, No 1, (1999); arXiv: gr-qc/9902148.

\bibitem{Wil}
C. M. Will,
``The Confrontation between General Relativity and Experiment,''
 Living Rev. Relativity, {\bf 9}, (2006), 3;
http://www.livingreviews.org/lrr-2006-3.

\bibitem{CFR}
G. Clement, J. C. Fabris and M. Rodriges, 
``Phantom black holes in Einstein-Maxwell-Dilaton theory,''  
Phys. Rev. D {\bf 79}, 064021 (2009);
arXiv: 0901.4543.

 \bibitem{ACFR}
 M. Azreg-A\"inou, G. Cl\'ement, J. C. Fabris and M. E. Rodrigues,
 ``Phantom Black Holes and Sigma Models,''
 Phys. Rev. D {\bf 83}: 124001 (12 pages) (2011).

 \bibitem{I-Sym-17} 
V. D. Ivashchuk,  ``On Brane Solutions with Intersection Rules Related to Lie Algebras'', Review,   Symmetry {\bf 9} (8), 155 (54 pages) (2017).

%\bibitem{AI}
%M.E. Abishev,  V. D. Ivashchuk, in preparation.



\end{thebibliography}
\end{document}